# CyRes – Avoiding Catastrophic Failure in Connected and Autonomous Vehicles

(Extended Abstract)

Carsten Maple
WMG, University of Warwick, Coventry, Warwickshire, United Kingdom,
cm@warwick.ac.uk

Peter Davies
Thales, United Kingdom,
peter.davies@thalesesecurity.com

Kerstin Eder
Department of Computer Science, University of Bristol, Bristol, United Kingdom,
kerstin.eder@bristol.ac.uk

Chris Hankin
Department of Computing, Imperial College London, London, United Kingdom,
c.hankin@imperial.ac.uk

Greg Chance
Department of Computer Science, University of Bristol, Bristol, United Kingdom,
greg.chance@bristol.ac.uk

Gregory Epiphaniou
WMG, University of Warwick, Coventry, Warwickshire, United Kingdom,
gregory.epiphaniou@warwick.ac.uk

**Abstract**

Existing approaches to cyber security and regulation in the automotive sector cannot achieve the quality of outcome necessary to ensure the safe mass deployment of advanced vehicle technologies and smart mobility systems. Without sustainable resilience hard-fought public trust will evaporate, derailing emerging global initiatives to improve the efficiency, safety and environmental impact of future transport. This paper introduces an operational cyber resilience methodology, CyRes, that is suitable for standardisation. The CyRes methodology itself is capable of being tested in court or by publicly appointed regulators. It is designed so that operators understand what evidence should be produced by it and are able to measure the quality of that evidence. The evidence produced is capable of being tested in court or by publicly appointed regulators. Thus, the real-world system to which the CyRes methodology has been applied is capable of operating at all times and in all places with a legally and socially acceptable value of negative consequence.

## 1. Introduction

The rapid digitisation of automotive vehicles and infrastructure, designed to realise the potential of Connected and Autonomous Vehicles (CAVs), has vastly increased the probability of catastrophic and unrecoverable failure[1]. Current and pending regulations and legal infrastructure, together with technical methodologies developed over the past 50 years, or borrowed from the IT sector, are contributing to the issue, rather than mitigating it. For example, existing maturity models initially developed to drive policy implementation and process assessment, fail to capture all resilience lines of effort adequately, due to the complexity and size of the vehicular cyberinfrastructure[2].

Over the long term, existing approaches to cyber security and regulation in the automotive sector cannot achieve the quality of outcome necessary to ensure the safe mass deployment of advanced vehicle technologies and smart mobility systems. Without sustainable resilience hard-fought public trust will evaporate, derailing emerging global initiatives to improve the efficiency, safety and environmental impact of future transport.

This paper introduces an operational methodology, suitable for standardisation, for which:

1. The methodology itself is capable of being tested in court or by publicly appointed regulators.
2. Operators understand what evidence should be produced by it and are able to measure the quality of that evidence.
3. The evidence produced is capable of being tested in court or by publicly appointed regulators.

Typically this will mean that the real-world system to which the methodology has been applied is capable of operating at all times and in all places with a legally and socially acceptable value of negative consequence.

## 2. Context

Evolution from traditional automobiles has resulted in automotive systems that have become **large-scale ad-hoc heterogeneous system of systems** with emergent and dynamic properties and for which the boundaries of ownership, control and responsibility – if not liability – are ill-defined. Whist the system has increased in complexity, individuals and organisations remain obligated to have robust engineering processes which must provide, to the standard the law expects, the type of evidence that is required when justifying the safety, privacy and market compliance of their components, systems and platforms.

Traditionally, post-sale responsibilities for vehicle manufacturers have been, principally, to provide warnings regarding newly discovered risks. However, the authors have demonstrated at least six classes of cyber threat that are inherent and exploitable in any complex cyber-physical system; in that context, and with that knowledge, the existing post-sale responsibilities represent a theoretically infeasible task. As the vision of connected and automated mobility (CAM) is realised, there is a need for organisations to move from concentrating solely on manufacturing, to delivering robust and resilient engineering practices from manufacture to operation.

---

[1] Maple, C. (2017). Security and privacy in the internet of things. Journal of Cyber Policy, 2(2), 155-184.
[2] Douglas Gettman et al. (2017). Guidelines for Applying Capability Maturity Model Analysis to Connected and Automated Vehicle Deployment, ITS Joint Programme Office, U.S. Department of Transport.

The nature of CAV systems means the cyber attack surface is, practically, infinite[3] with the point of specific ingress changing instantly, and a potentially very high rate of propagation within and between mobility elements. That renders it impossible to simplify cyber incident analysis with existing network model taxonomy descriptions[4]. Further, defending one attack may, in fact, guarantee the success of another, in part due to the limited collective security measures in place. So, for example, 'rebooting' a large vehicle fleet in operation would be infeasible and most likely undesirable in any case; but were it achieved it would undoubtedly, in this less than fully-analysed state, serve as a vector for ingress of malware. Traditional engineering approaches, such as the V-model, are incapable of addressing such challenges, and new methods should be developed to ensure the cyber resilience, cyber security and cyber safety of advanced vehicles and infrastructure.

## 3. Cyber Resilience Principles

The authors have held a number of workshops and focus groups over the last two years, to identify the key challenges and to develop principles and methods to overcome these. This has resulted in three Principles:

- Increase the probability of *detection, understanding* and *acting on* cyber events;
- Increase the number of *Engineered Significant Differences*; and
- Invoke a continuum of *Proactive Updates*.

To assess the efficacy of the engineering approach and to provide a basis of evidence-based certification, we propose the following certification arguments:

- Probability of detecting threats
- Probability of understanding threats
- Rate of deploying mitigating actions
- Time for a threat to propagate
- Quantity of Engineered Differences
- Frequency of Proactive Updates

## 4. Measuring Resilience

The mission performance metric, P(t), measuring the total performance of a system over time can be defined to produce a value between 0 and 1, where P(t) = 0 corresponds to the system not functioning at all and P(t) = 1 corresponds to a fully functioning system. In our case, we can define the mission performance metric P(t) as the average performance metric across all vehicles.

We can use this performance metric to assess resilience during an incident. We define an event, *E*, to occur over an interval of time *t* = [*0, T*], where time *t* = *0* represents the start of the incident, and *t* = *T* is the end. The start of an event is identified when for at least one vehicle the performance metric is less than 1. The end of an event is determined when P(t) =

---

[3] The Automotive system will include Satellites, 3G/4G/5G, WiFi, DSRC, CV2X, Bluetooth, GNSS, USB, OBD into CANbus, and a range of other possible connections, including access to IoT networks, chipset IP, training data sets.
[4] Zeinab El-Rewini et al. (2020). Cybersecurity challenges in vehicular communications. Journal of Vehicular Communications, 23, p.100214

1, that is the system has fully recovered, or some value $P_A$, where $P_A$ is a level deemed acceptable by regulation. Figure 2 illustrates an example of a performance metric during an event.

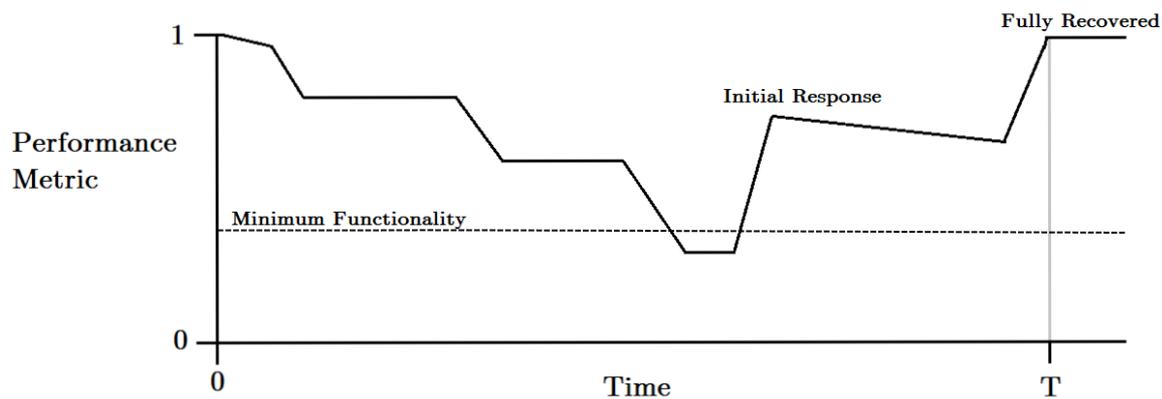

Figure 2: System performance over an event

In this example, the performance decreases in stages. For a short period of time t=to, it falls below a threshold which is considered to be the minimum performance required for the system to be useful. To limit the amount of time the system spends in this state, the manufacturer issues an initial response which provides a partial recovery. Towards the end of the event, the manufacturer issues a permanent fix and the system returns to being fully operational.

The length of time between the start of the event and the first time it decreases system performance below the minimum functionality threshold can be termed the time-to-failure, and the period after this until the performance returns to an acceptable level is termed the time-to-recovery. For a particular event, the time interval [0,T] can be partitioned into the time it takes for the manufacturer to: detect the incident; understand the cause; and respond to it. We can then define the resilience to be the integral of the performance metric during the event.

If no response is made to an incident, then the system may reach a point of catastrophe at time $t = t_c$, the time for a threat to propagate. This could be defined in multiple ways, for instance, it could be seen as the point where it is no longer financially viable to provide a fix or where $P(t)$ decreases below the minimum functionality threshold for a certain length of time. For some smaller incidents, this may never happen in which case $t_c = \infty$.

For complete cyber resilience, a significant response must have been made before $t = t_c$ otherwise catastrophe will occur. Assuming the main response is provided close to time $t = T$ at the end of the response stage, we say the system is completely cyber resilient if $T < t_c$ and this can be consistently achieved for all possible events.

If for a particular event $T < t_c$ is not achieved then we may be in one of three possible cases: 1) we have failed to detect the event in time; 2) we have not understood the problem, and 3) we have understood the problem, but we have not had enough time to develop a fix and release it.

Complete cyber resilience might be achievable for closed systems, but for autonomous vehicles this is unrealistic. Instead we view the cybersecurity challenge with the aim of reducing the probability a random event leads to catastrophe to a level which can jointly be covered by insurance and acceptance of a certain amount of risk. In our mathematical model, we need to reduce the amount of loss. There are three ways of doing this:

- by detecting, understanding and responding quicker;
- by increasing the performance metric during the event, by improving or maintaining the performance metric for smaller groups of vehicles; or
- by increasing the frequency of updates, thus ensuring the attacker needs to amend the attack to achieve the attack goal.

## 5. CyRes – A dynamic system methodology for cyber resilient CAVs

CyRes, an overview of which is presented in Figure 1, presents a whole-life methodology focussed on operation but also covering design, manufacture, deployment and redevelopment. At the design stage, the standard systems engineering 'V-model' with all its concomitant certainty may be used for subsystems only where those systems are bounded, known and predictable. Where this is not the case the design objective prior to launch must be to achieve coverage by instrumenting the system so that it can be more easily monitored, understood and adapted in operation. This ensures that it can operate effectively at the rate the system metamorphoses in the face of emergent properties and threats. The rate of metamorphosis will not permit continuous repetition of a conventional full life-cycle.

During the operation phase, monitoring the system allows threats to be detected. Once a threat has been detected, it first needs to be understood, using on-board diagnostics, potentially supported by off-board simulation, to assess the impact of the threat on the system. Candidate responses can then be simulated and, as appropriate, deployed. Through the use of a Central Intelligence function, new threats, optimal outcomes, and assessment of the mitigations can be stored. This knowledge can be used to expedite the process in the future, supporting rapid retrieval of known threats, scenarios for suggested remediation and the efficacy of these.

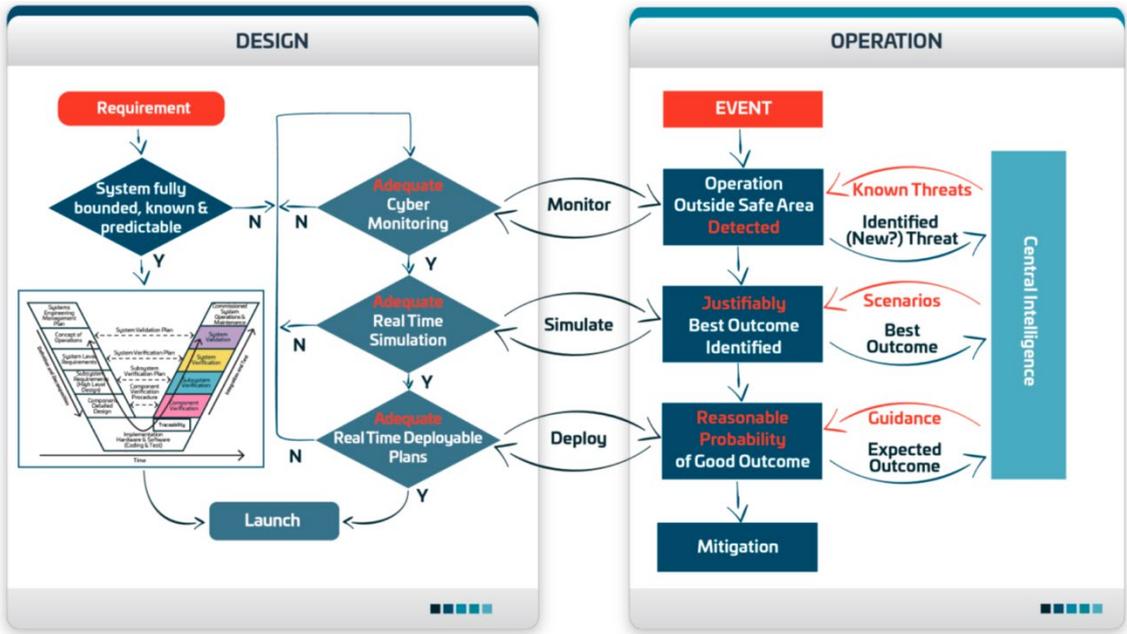

Figure 1: The CyRes methodology

CyRes employs real time cyber monitoring to detect abnormal events occurring outside the design limits. Theorem proving can be used to classify whether events are within design limits or not. Detection of a threat may occur in resource-rich environments before the event, or in resource-rich or resource-constrained environments after the event. Detection of a threat post-event is usually through signature-based, anomaly-based or specification-based methods. Through monitoring the system, the propagation of known threats can be examined.

Having detected threats, on-board diagnostics supported by simulation should then be undertaken to determine how close to the design limit future operating states might become. If a cyber-attack on a vehicle is detected, it is critical to understand the impact the attack may have on the vehicle and its environment in order to take appropriate action to contain and mitigate the attack. It is particularly important to know whether action needs to be taken immediately - whether it is necessary to bring the vehicle to the nearest garage on the day or within a week - or whether it is sufficient to raise this issue during planned maintenance. On-board diagnostics and simulation provide this understanding. Conditions that would result in failure, and their likelihood of occurring, as well as the optimal remediation strategy for the threat, should also be found through simulation.

Deployment involves the time-critical strategic execution of containment and mitigation plans. The system is monitored to ascertain whether it is responding as expected. In cases where there is an undesirable response, the deployment should be withdrawn, and an immutable log of all decisions, actions and supporting evidence be updated. DevOps practices and container technologies (including orchestration) allow unprecedented rates of system change while maintaining control and oversight of the operational parameters of the system. Uber employs such an approach, launching more than a million containerised batch jobs per day, building 5,000 software items per day and deploying thousands of microservices across 35 clusters[5].

In electro-mechanical systems, each system and platform is different and consequently each is expected to be susceptible to failure in a different way and at a different time. As such, the potential harm arising from failure occurs with statistical probability, one device at a time; this principle forms the basis of standard safety calculations. In digital systems, an identified fault could manifest in all digitally identical systems at the same time, thereby giving rise to global catastrophic failure. That is, at the level of the overall system, rather than at the individual device. Functionality defence by heterogeneity, inspired by the biological phenomenon of the human race surviving deadly viruses because of the diversity arising from heterogeneity, has been proposed as a paradigm for securing systems[6]. This inspires the concept of *engineered significant differences*, the deliberate introduction of significant differences between systems and platforms from design time to manufacture and operation. These differences are imperceptible to the user or operator but may prevent all systems being identically affected by cyber-attacks.

---

[5] Y. Liu, "Only slightly bent - Uber's Kubernetes Migration Journey for microservices," [Online]. Available: https://kccncna19.sched.com/event/Uabh/only- slightly-bent-ubers-kubernetes-migration-journey-for-microservices-yunpeng-liu- uber.

[6] Sharman, R., Rao, H.R., Upadhyaya, S., Khot, P., Manocha, S. and Ganguly, S., 2004, January. Functionality defense by heterogeneity: a new paradigm for securing systems. In 37th Annual Hawaii International Conference on System Sciences, 2004. Proceedings of the (pp. 10-pp). IEEE.

Proactive updates are continuous software updates which may be applied in an attempt to defend against evolving cyber-attacks. These can be generated through the simulation of the systems incorporating known and anticipated attacks. Currently, software vendors attempt to distribute security patches in advance of vulnerability disclosure to mitigate risks before they become issues. CyRes builds on this common practice and applies it to CAVs, extended through large-scale simulation, formal methods and distributed threat monitoring to derive a cyber course of action (COA) to respond to incidents. Proactive updates should exploit engineered significant difference, and by introducing controlled software variability, malware contagion can be contained. Since software updates represent a threat in their own right, updates must be assured to come from a trusted source and have undergone suitable validation and verification before being deployed.

## 6. Conclusion

We have demonstrated the need for a fundamental change in engineering practices for such systems and commend the CyRes approach as a new methodology to address this.